\documentstyle[sprocl]{article}
\input{psfig}
\def\be{\begin{equation}}
\def\ee{\end{equation}}
\def\bea{\begin{eqnarray}}
\def\eea{\end{eqnarray}}

\begin{document}
\title{
NEW METHODS FOR EXTRACTING THE CKM ANGLE $\gamma$
USING $B^\pm\to D^0 K^-$; $\overline D^0 K^-$
}
\author{David Atwood }
\address{
Thomas Jefferson National Accelerator Facility, 12000 Jefferson Ave.
Newport News, VA 23606, USA
}
%
%
%
%
%
%
%
%
\maketitle\abstracts
{
In this talk I will discuss the extraction of the CKM angle $\gamma$ at
$B$-factories through the interference of the subprocesses $B^-\to K^- D^0$ and
$B^-\to K^-\overline D^0$. This seemingly impossible interference may be
accomplished by allowing both $D^0$ and $\overline D^0$ to decay to a common
final state. If only CP eigenstate decay modes of the $D$ are considered, the
branching ratio for $B^-\to \overline D^0 K^-$ must be experimentally
determined in order to extract $\gamma$. 
I describe why
this determination is
likely to be experimentally impossible. On the other hand, if more general  $D$
decays are considered, the angle $\gamma$ may then be determined. In fact, it
is possible that a reasonable determination of $\gamma$ may be made with
$O(10^8)$\ $B$'s.
\vspace*{.25 in}
\begin{center}
{\it Contributed to Proceedings of {\bf B Physics and CP Violation},
Honolulu, Hawaii, March 1997.}
\end{center}
}

\section{Introduction}

The Unitarity of the Cabibbo-Kobayashi-Maskawa (CKM) matrix is one of the
fundamental predictions of the Standard Model~\cite{ckmref}. 
The experimental activity at the
various $B$ factories will therefore concern itself largely with the accurate
determination of the various elements of this matrix. While the magnitude of
CKM elements may be determined by the rates of appropriately chosen processes,
direct measurement of the phases of these elements require the observation of
CP violation. Indeed, with the exception of strong CP violation, the phase in
the CKM matrix is the only place the Standard Model admits CP violation.
Indeed there are good reasons to believe that CP violation will be present at
an observable level in the $B$ system 
since
the CKM matrix in three
generations will in general have a complex phase 
and
CP violation has been known for 30
years in the $K_L$ system 
which may be explained though such a phase.

This being the case, the standard model makes makes definite prediction
regarding CP violation that will be present in B physics. This may be
summarized with the usual ``Unitarity Triangle''~\cite{quinpdb}
which shows how 
how the CP violation from the CKM matrix will be distributed 
among different
$B$ decay channels. 
In this talk I will focus on the determination of the angle 
$\gamma$ which is the phase of the element $V_{ub}$ in the CKM 
parameterization 
of~[~\cite{wolf}~].

\section{Determining $\gamma$ with $B^-\to D^0 K^-$}

On the quark level, the determination of $\gamma$ which I will discuss is
based on the interference of the tree level decays $b\to c \overline u s$ with
$b\to u \overline c s$ (and their charge conjugates). The CKM phase difference
between these two amplitudes is readily seen to be $\gamma$ however it is not
obvious how these channels, leading to seemingly different final states, can
have any quantum mechanical interference.

It is only by considering certain specific hadronic final states common to both
sub-processes that the desired interference may be obtained. First we specify
that $b\to c \overline u s$ hadronize as $B^-\to D^0 K^-$ and that $b\to u
\overline c s$ hadronize as $B^-\to \overline D^0 K^-$. These may interfere
only if both the $D^0$ and $\overline D^0$ decay to a common final state $X$.
In what follows, we will consider what choices of $X$ can lead to a practical
method for the determination of $\gamma$.

This clever method of extracting $\gamma$ was first proposed in 
1990~\cite{glw}
and has
since been studied extensively in the case where $X$ is a CP 
eigenstate~\cite{glwstudied} 
(which we will refer to as the Gronau-London-Wyler (GLW) method)
and
also for more general values of $X$~\cite{izi,ads}. 
Recently it has been realized~\cite{ads} that the
recipe for extracting $\gamma$ 
as originally proposed using
CP eigenstate modes require the
knowledge of $Br(B^-\to \overline D^0 K^-)$
and that this
is virtually impossible to
obtain experimentally. In this talk I will emphasis that considering more
general values of $X$ which are not CP eigenstates allows one to get around
this difficulty and develope a practical method for the determination of
$\gamma$.

Indeed, the only other method for determining $\gamma$ is through oscillation
effects in the $B_s$~\cite{quinpdb}. 
Methods based on the interference of $b\to u\overline c
s$ with $b\to c\overline u s$ are thus of great importance since they may be
used at $\Upsilon(4s)$ $B$-factories which do not produce $B_s$ mesons

\section{Using CP-eigenstate $D^0$ Decays}

Let us first consider the case where $X$ is a CP eigenstate. Some examples of
such a decay are 
$D^0\to \pi^+\pi^-$, 
$D^0\to K_s\pi^0$, 
$D^0\to K_s\eta$ etc.  
For instance if $X=K_s\pi^0$ the $b\to c$ channel hadronizes as 
$B^-\to K^- (D^0\to K_s\pi^0)$ 
while the 
$b\to u$ transition hadronizes as
$B^-\to K^- (\overline D^0\to K_s\pi^0)$. Overall both processes lead to a
common final state (ie. $K^-K_s\pi^0$ with $M_{K_s\pi^0}=M_{D^0}$) and they
will interfere. In general, two phases will enter into the interference
process, the CP odd phase $\gamma$ which we wish to measure and the
rescattering phase $\xi$
resulting from the fact that the final state 
$K^- D^0$ 
rescattering phase
is different from that of 
$K^- \overline D^0$.

Let us define 
\begin{eqnarray}
&
a(K)=Br(B^-\to K^- D^0)
\ \ \ b(K)=Br(B^-\to K^- \overline D^0) \ \ \
c(X)=Br(D^0\to X)
&
\nonumber\\
&
c(\overline X)=Br(D^0\to \overline X)
\ \ \ d(K,X)=Br(B^-\to K^- [X]) \ \ \
\overline d(K,X)=Br(B^+\to K^+ [\overline X])
&
\nonumber
\end{eqnarray}
where $[X]$ indicates that it proceeds through the interfering $D^0$ and
$\overline D^0$ channels.
Then $d$ and $\overline d$ are given by:
\begin{eqnarray}
d(K,X)&=&
a(K)c(X)+b(K)c(\overline X)
+2\sqrt{a(K)b(K)c(X)c(\overline X)}\cos(\xi+\gamma)
\nonumber\\
\overline d(K,X)&=&
a(K)c(X)+b(K)c(\overline X)
+2\sqrt{a(K)b(K)c(X)c(\overline X)}\cos(\xi-\gamma)
\nonumber
\end{eqnarray}

Let us now assume that $a(K)$, $b(K)$, $c(X)$ and $c(\overline X)$ are known
experimentally (if $X$ is CP eigenstate $c(X)=c(\overline X)$). Then if
$d(K,X)$ and $\overline d(K,X)$ are measured, the two equations above
may be solved for the two unknown phases $\xi$ and $\gamma$. This is the
essence of the GLW method. The assumption that $b(K)=Br(B^-\to K^- \overline
D^0)$ can be measured however requires careful scrutiny since it may be
estimated~\cite{ads} that $b(k)\approx 3\times 10^{-6}$ is rather small.

In order to measure $b(K)$ we need some way to tag the $\overline D^0$ and in
particular to tell it from a $D^0$. Logically, the are two possible ways one
can accomplish this tagging, via a hadronic mode (which is the method
considered in the literature to date~\cite{glwstudied})
or via a semi-leptonic mode.

Possible hadronic modes which tag $\overline D^0$ are decays of $\overline D^0$
which are Cabibbo allowed. For instance the decay $\overline D^0 \to K^+\pi^-$
where $Br(\overline D^0\to K^+\pi^-)=3\times 10^{-2}$~[~\cite{pdb}~]. 
The total decay rate for
the chain $B^-\to K^- [\overline D^0\to K^+\pi^-]$ will thus be $\sim 10^{-7}$.
Unfortunately $D^0$ may also decay into $K^+\pi^-$ although this decay is
doubly Cabibbo suppressed; in particular $Br(D^0\to K^+\pi^-)\approx 3\times
10^{-4}$~[~\cite{pdb}~]. 
The primary decay $B^-\to K^- D^0$ 
however
has a branching ratio of $\sim
3\times 10^{-2}$ so that chain $B^-\to K^- [\overline D^0\to K^+\pi^-]$ is also
$\sim 10^{-7}$. Since both chains lead to the same final state, there will be
$\sim 100\%$ interference effects between the two channels and so $b(K)$ cannot
be determined in isolation. All possible hadronic tags of $\overline D^0$ will
be likewise afflicted with these interference effects.

A semi-leptonic tag is any decay of the form $\overline D^0\to e^-\overline\nu
X_s$. 
This signature however is subject to a background from
$B^-\to
e^- \overline\nu X$ 
which is $10^6$ times larger.

\section{Using Non-CP Eigenstates}

The key to extracting $\gamma$ without the use of $b(K)$ is to take advantage
of precisely the large interference effects which prevented the determination
of $b(K)$ above. Consider the case where $X$ is not a CP eigenstate in
particular where $D^0\to X$ is doubly Cabibbo suppressed, for example
$X=K^+\pi^-$. Now the strong phase difference between the decay chain $B^-\to
K^- [D^0\to X]$ and $B^-\to K^- [\overline D^0\to X]$ is $\zeta=\xi+\eta$ where
$\xi$ is the strong phase difference arising from the rescattering of $D^0 K^-$
versus $\overline D^0 K^-$ and $\xi$ arises from the phase difference between
$D^0\to X$ versus $\overline D^0\to X$.

In general each possible instance of $X$ will have a different
value of $\xi$ so that if two choices of $X$ are used, the set of equations
above are replaced by the system of four equations:
\begin{eqnarray}
d(K,X_i)&=&
a(K)c(X_i)+b(K)c(\overline X_i)
+2\sqrt{a(K)b(K)c(X_i)c(\overline X_i)}\cos(\zeta_i+\gamma)
\nonumber\\
\overline d(K,X_i)&=&
a(K)c(X_i)+b(K)c(\overline X_i)
+2\sqrt{a(K)b(K)c(X_i)c(\overline X_i)}\cos(\zeta_i-\gamma)
\nonumber
\end{eqnarray}
for $i=1,2$.
Assuming that $a(K)$, $c(X_i)$ and $c(\overline X_i)$ are already known and
$d(K,X_i)$ and $\overline d(K,X_i)$ are then measured, 
the system above provides four equations for the four unknowns
$\{\gamma$, $\zeta_1$, $\zeta_2$, $b(K)\}$ which can in principle be solved
(and as a by product we also get the value of $b(K)$). 
These equations will be non-degenerate if either
$c(X_1)/\overline c(X_1) \neq c(X_2)/\overline c(X_2)$
or $\zeta_1\neq\zeta_2$ which will occur if 
both $X_1$ and $X_2$ are not 
CP eigenstates or if $X_1$ is a CP eigenstate and $X_2$ is not.

\section{Improvements} 

Note that the system of equations above are quartic in nature and so in
addition to the ambiguity between $\gamma$ and $-\gamma$, there is a four-fold
ambiguity in the determination of $\gamma$. If, however, a third state is also
used, the resulting system of six equations is over-determined and the
four-fold ambiguity is resolved.

Indeed there are 
a number of possible non CP-eigenstate modes that may be used for 
$X$, in particular $K^+\pi^-$, $K^+\rho^-$, $K^+ a_1^-$, $K^{*+}\pi^-$ etc.
One may further generalize to related $B^-$ decays such as 
$B^-\to K^{*-} D^0$ and $B^-\to K^- D^{*0}$ all of which tend to build up the
statistics for the determination of $\gamma$.

The accuracy in determining $\gamma$ through this method depends on the 
value of $\gamma$ as well as the completely unknown values of the strong phase
shifts involved. This error will typically~\cite{ads} be between $5^\circ$ and
$20^\circ$ given the total number of $\Upsilon(4s)$ of $\sim 10^8$ (not
including acceptance factors).

Finally, one can improve the determination of $\gamma$ by considering 3-body
decays of $D^0$. In particular if $D^0\to K^+\pi^-\pi^0$ additional information
may be obtained by considering the distribution as a function of the energy of
the $K^+$ and $\pi^-$ in the rest frame of the $D^0$.

\section{Conclusion}

In conclusion, we have see that the original method of GLW for determining
$\gamma$ has a problem due to the fact that interference prevent the
determination of $Br(B^-\to \overline D^0 K^-)$ through hadronic decays of
$\overline D^0$. We can, however, exploit these effects to salvage a method for
determining $\gamma$ and, since these interferences are between two roughly
equal amplitudes, CP violating effects will be $O(100\%)$!. Assuming that modes
such as $K^-+n\pi$ may be tagged with a reasonable efficiency, there is a
prospect that the luminosities typical of B factories will give a determination
of $\gamma$ to a precision of $5-20^\circ$ where the exact precision obtainable
depends on unknown strong rescattering phases. If this is achieved, it could
have a significant impact on the determination of CKM parameters and more
importantly, it is probably the only way of directly determining $\gamma$ at
$\Upsilon(4s)$ $B$-factory experiments.

\vspace*{.1 in}

This research was supported in part by the U.S. DOE contract
DC-AC05-84ER40150.


\section*{References}

\begin{thebibliography}{99}


\bibitem{ckmref}
M.~Kobayashi and T.~Maskawa, Prog. Theor. Phys. 
49, 652 (1973).


\bibitem{quinpdb}
For a review see, e.g. the article of
H.~Quinn in {\it Particle Data Book}, R.M. Barnett et al., Phys. Rev.  
D54, 507 (1996).

\bibitem{wolf}
L. Wolfenstein, Phys. Rev. Lett. 51, 1945 (1984).

\bibitem{glw}
M. Gronau and D. London, Phys. Lett. B253, 483 (1991);
M. Gronau and D. Wyler, Phys. Lett. B265, 172 (1991); see also
I.I.Y. Bigi and A.I. Sanda, Phys. Lett. 211B, 213 (1988) where similar
ideas are discussed.


\bibitem{glwstudied}
Several feasibility studies have been conducted.  See, for instance,
M. Witherell, private communication;
S. Stone, Nucl. Instrum. Meth. A 333, 15 (1993);
A. Snyder, BaBar notes \# 80, \#84;
I. Dunietz, Z. Phys. C56, 129 (1992);
I. Dunietz, in B Decays, 2nd Edition, S. Stone ed.
(World Scientific, Singapore, 1994), p. 550.


\bibitem{izi}
I. Dunietz, Phys. Lett. B270, 75 (1991).

\bibitem{ads}
D.~Atwood, A.~Soni and I.~Dunietz, Phys. Rev. Lett. 78, 3257 (1997);
D. Atwood, I. Dunietz, and A. Soni, in preparation.


\bibitem{pdb}
R.M. Barnett et al. (Particle Data Group), Phys. Rev.  
D54, 1 (1996).






%
%
%
\end{thebibliography}
\end{document}